\documentclass[sigconf,natbib=true,anonymous=false]{acmart}

\renewcommand\footnotetextcopyrightpermission[1]{} 
\usepackage{enumitem}
\usepackage{multirow}
\usepackage{float}

\usepackage{hyperref} 
\usepackage{cleveref} 
\hypersetup{linkcolor = black, 
	    urlcolor = blue,   
            citecolor = blue} 

\usepackage{subfigure}

\AtBeginDocument{%
  \providecommand\BibTeX{{%
    \normalfont B\kern-0.5em{\scshape i\kern-0.25em b}\kern-0.8em\TeX}}}

\acmISBN{978-1-4503-XXXX-X/18/06}

\begin{document}

\title{End-to-end training of Multimodal Model and ranking Model}

\author{Xiuqi Deng, Lu Xu, Xiyao Li, Jinkai Yu, Erpeng Xue, Zhongyuan Wang, Di Zhang,\\ Zhaojie Liu, Yang Song, Guorui Zhou, Na Mou, Shen Jiang} 
\orcid{1234-5678-9012}
\affiliation{%
  \institution{KuaiShou Inc.}
  \city{Beijing}
  \country{China}
}
\renewcommand{\shortauthors}{Deng, et al.}

\begin{abstract}
Traditional recommender systems heavily rely on ID features, which often encounter challenges related to cold-start and generalization. Modeling pre-extracted content features can mitigate these issues, but is still a suboptimal solution due to the discrepancies between training tasks and model parameters. End-to-end training presents a promising solution for these problems, yet most of the existing works mainly focus on retrieval models, leaving the multimodal techniques under-utilized. In this paper, we propose an industrial multimodal recommendation framework named EM3: End-to-end training of Multimodal Model and ranking Model, which sufficiently utilizes multimodal information and allows personalized ranking tasks to directly train the core modules in the multimodal model to obtain more task-oriented content features, without overburdening resource consumption. First, we propose Fusion-Q-Former, which consists of transformers and a set of trainable queries, to fuse different modalities and generate fixed-length and robust multimodal embeddings. Second, in our sequential modeling for user content interest, we utilize Low-Rank Adaptation technique to alleviate the conflict between huge resource consumption and long sequence length. Third, we propose a novel Content-ID-Contrastive learning task to complement the advantages of content and ID by aligning them with each other, obtaining more task-oriented content embeddings and more generalized ID embeddings. In experiments, we implement EM3 on different ranking models in two scenario, achieving significant improvements in both offline evaluation and online A/B test, verifying the generalizability of our method. Ablation studies and visualization are also performed. Furthermore, we also conduct experiments on two public datasets to show that our proposed method outperforms the state-of-the-art methods.

\end{abstract}




\maketitle
\pagestyle{plain}


\section{Introduction}
For the past decades, recommender systems (RS) have achieved a fabulous performance\cite{zhang2019deep}.
Large-scale industrial RS usually use unique identities (ID) to represent users and items. Benefited from its strong abilities to remember and capture the user-item relationships, this ID-based paradigm dominates the RS fields until now \cite{yuan2023go, zhao2023embedding}. 

But there are still some shortcomings. On the one hand, ID embeddings have the cold-start problems because of data sparsity \cite{he2016vbpr}. On the other hand, it may cause the estimation variance on items with similar materials \cite{chen2021efficient}, known as the generalization problems. One solution is to model content features so that inference can be made without interaction records \cite{geng2022recommendation}, which shows better accuracy than general ID-based models \cite{zhou2023tale}. This can be done in two ways: pre-extraction (PE) and end-to-end (E2E) \cite{zhou2023comprehensive}.

The PE paradigm extracts frozen features from content models and feeds them into recommendation models (RM) as common features, also known as the two-stage paradigm \cite{yuan2023go, mo2015image, zhou2023bootstrap}. However, the content-oriented pre-training tasks do not match well with the downstream personalized task \cite{xu2021e2e, liu2023megcf}. Besides, industrial RM require continuous training to follow the time-varying online distribution \cite{liu2020category}, while the content models remain frozen, giving rise to parametric mismatch. Both of the above reasons will result in suboptimal performance.

On the contrary, E2E refers to feeding low-level content features into RM and updating the content model together. Earlier studies have shown that E2E performs better than PE \cite{yuan2023go, liu2020category}. However, most of them are based on two-tower \cite{he2016sherlock, tautkute2019deepstyle, elsayed2022end, yang2022gram, zhou2023tale} or session-based models \cite{xiao2022training, zhang2023multimodal, zhang2023language}, which are generally used as retrieval models. Few related studies on industrial ranking models only consider a single visual modality \cite{liu2020category, chen2016deep, ge2018image, chen2022hybrid}. Nowadays, online platforms usually have diverse content domains \cite{cui2022m6}, utilizing multimodal information can capture useful information that is invisible in the single modality and can handle the modality missing problem to get a suitable content representation \cite{zhou2023comprehensive}. To summarize, the E2E training of multimodal model and ranking model is a valuable and promising direction that has not been fully explored.

\begin{figure*}[htb]
  \centering
  \includegraphics[width=\linewidth]{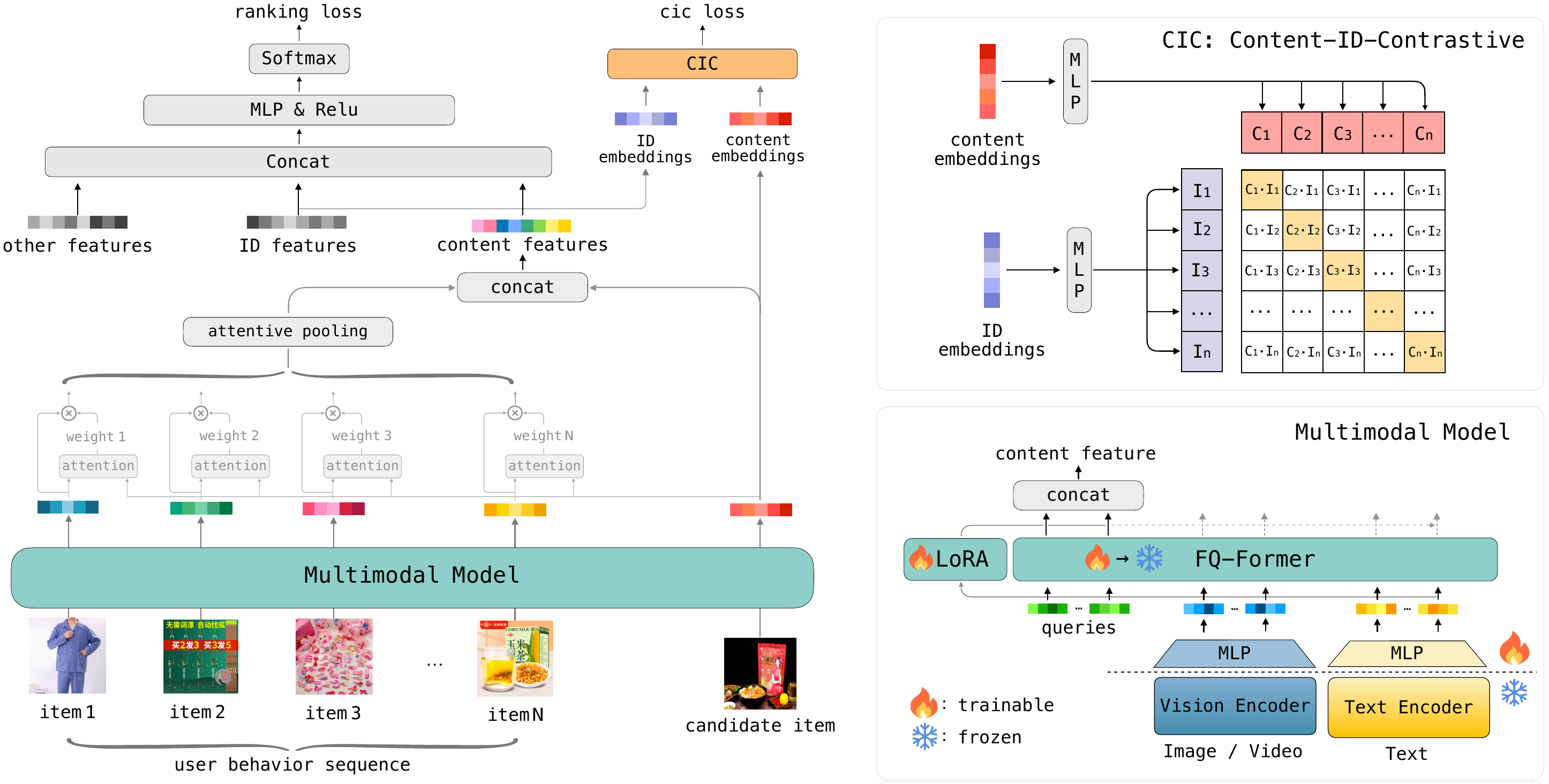}
  \caption{
     The overall framework of EM3. 
     (a) Multimodal Model: aims to fuse modalities and shares parameters across different items. 
     (b) CIC: a self-supervised task to align content and ID by maximizing the cosine similarities of the pairs on the diagonal.
  }
  \label{fig:framework}
  \Description{}
\end{figure*}

In this paper, we propose an industrial multimodal recommendation framework named \textbf{EM3}: \textbf{E}nd-to-end training of \textbf{M}ultimodal \textbf{M}odel and ranking \textbf{M}odel. As shown in \cref{fig:framework}, EM3 sufficiently utilizes multimodal information and allows personalized ranking tasks to directly train the core modules in multimodal model, obtaining more task-oriented content representations without overburdening resource consumption. First, inspired from BLIP2 \cite{li2023blip}, we propose Fusion-Q-Former, which consists of transformers and a set of trainable queries, to fuse different modalities and generate fixed-length and robust multimodal embeddings. Second, in our sequential modeling for user content interest, we utilize Low-Rank Adaptation technique \cite{hu2021lora} to alleviate the conflict between huge resource consumption and long sequence length. Moreover, we propose a novel Content-ID Contrastive learning task to complement the advantages of content and ID by aligning them with each other, obtaining more task-oriented content embeddings and more generalized ID embeddings. In our comprehensive experiments, we verify the effect of EM3 on two different ranking models in our system, achieving significant improvements on the offline dataset with billions of records and online A/B test, contributing to millions of revenue. A series of ablation studies and a visualized analysis are also presented. Furthermore, we conduct experiments on two public datasets to show that our proposed method outperforms the state-of-the-art methods in terms of recommendation accuracy, and the source code is available at \url{https://github.com/em3e2e-anonymous/em3}.

Our main contributions can be summarized as follows:
\begin{enumerate}[leftmargin=20px]
\renewcommand{\labelenumi}{\roman{enumi}.}
    \item To the best of our knowledge, this is the first work to propose an industrial framework for E2E training of multimodal model and ranking model, verifying the value and feasibility of this direction in both academia and industry.
    \item We propose Fusion-Q-Former to fuse different modalities, which consists of transformers and a set of trainable queries, generating fixed-length and robust multimodal embeddings.
    \item We utilize Low-Rank Adaptation technique to alleviate the conflict between the huge number of trainable parameters and the sequence length in sequential modeling.
    \item We propose a novel Content-ID-Contrastive learning task to complement the advantages of content and ID by aligning them with each other, obtaining more task-oriented content embeddings and more generalized ID embeddings.
\end{enumerate}

\section{Related Work}
Our work is closely related to three research areas: recommendation system, multimodal model, and multimodal recommendation.

\subsection{Recommender Systems}
Industrial RS are generally divided into two stages: retrieval and ranking. Retrieval narrows down the selection to a small set of items, while ranking provides more precise estimation \cite{zhao2023embedding}. In this paper, we focus on ranking models, which are usually single-tower and sensitive to resource consumption. 

In early ages, ranking model typically adopt collaborative filtering, logistic regression or matrix factorization \cite{sarwar2001item, mcmahan2013ad, koren2009matrix} to capture user-item relationships. Later, the contextual features, user profiles and item attributes are integrated into the RS through more sophisticated models \cite{geng2022recommendation} such as FM \cite{rendle2010factorization}, FFM \cite{juan2016field}, Wide\&Deep \cite{cheng2016wide} and DeepFM \cite{guo2017deepfm}. Recently, user interests are further excavated by sequential modeling such as DIN \cite{zhou2018deep} and SIM \cite{pi2020search}. Today, industrial ranking models are usually hybrid models that include the above various technologies.

\subsection{Multimodal Model}

The research on multimodal model was originally divided into two separate fields: 

\begin{itemize}[leftmargin=15px]
    \renewcommand{\labelenumi}{\roman{enumi}.}
    \item Computer Vision (CV) includes tasks such as image classification and object detection. Initially, CNN were widely used \cite{krizhevsky2012imagenet, simonyan2014very, szegedy2015going}. Recently, Vision Transformer (ViT) has achieved remarkable results \cite{dosovitskiy2020image, wang2021pyramid, liu2021swin}.
    \item Natural Language Processing (NLP) includes tasks such as machine translation and Q\&A. RNN \cite{sutskever2014sequence, bahdanau2014neural} and BERT \cite{vaswani2017attention, devlin2018bert, liu2019roberta, hu2021lora} have successively dominated for many years. Currently, large language models (LLM) are ushering in a new era of generative models \cite{radford2018improving, ouyang2022training}.
\end{itemize}

Nowadays, multimodal learning unifies these two domains, aiming to extract and understand multi-media information better when various modalities are engaged \cite{jabeen2023review}. Some researchers focus on fusing different modalities into a single embedding, such as the single-flow paradigm in ViLT \cite{kim2021vilt} and two-flow paradigm in ViLBert \cite{lu2019vilbert}. Others maintain two independent representations by aligning them across domains, such as CLIP \cite{radford2021learning}. ALBEF \cite{li2021align} and BLIP \cite{li2022blip} integrate the two paradigms into a unified framework. BLIP2 \cite{li2023blip} further merge visual features into LLM via Q-Former, becoming one of the most representative approaches of Multimodal LLM (MLLM).

\subsection{Multimodal Recommendation}
Most of the previous works on multimodal recommendation are in the PE paradigm \cite{he2016vbpr, zhou2023comprehensive, mo2015image, zhou2023tale, zhang2021mining}, few E2E works are mainly based on two-tower models \cite{he2016sherlock, tautkute2019deepstyle, yang2022gram, zhou2023tale, zhang2021mining, tao2022self}, markedly different from the industrial ranking models, so we do not go into the PE or two-tower paradigm in detail in this section.

In the field of E2E training of multimodal model and ranking model, DeepCTR \cite{chen2016deep} incorporates a trainable CNN into the ranking model to capture visual features associated with advertising (ad). To accelerate the training speed, it groups samples with the same image into the same batch, thereby reducing the times of CNN forward propagation. CSCNN \cite{liu2020category} makes use of the plug-in attention module to feed e-commerce (e-com) categories as side information. This approach helps CNN extract diverse visual features. DICM \cite{ge2018image} not only adds a trainable image encoder on the item side but also implements it on user behavioral sequences, leading to a huge improvement. To enhance the training efficiency, it designs the AMS framework to separately deploy ID parts and multimodal parts on different devices. Besides, it selects the trainable-fixed hybrid paradigm by only updating the top fully connected (FC) layers of visual encoder. HCCM \cite{chen2022hybrid} combines the key benefits of the aforementioned methods: it utilizes ad categories as side information and extends its application to user behavioral sequences. 

In summary, the pioneering works have preliminarily realized E2E training in industrial ranking scenario, but there is still room for improvement: First, they only utilize single visual modality, while multimodal information in the current online environment has not been utilized. Second, despite they indicate the improvement of sequential modeling, the huge resource consumption will limit the sequence length in practice. Third, the complementary advantages between content and ID can be further leveraged. Last but not least, they all conduct experiments on batch-training CTR models, we can generalize it to a wider range of scenarios such as online-learning models and CVR models.

\section{Method}
\subsection{Ranking Model}
\label{chap:ranking model}
In the ranking model, there are a lot of features such as ID, context and session features. We concatenate and feed them into DNN, and the softmaxed output of DNN is used to optimize the negative log-likelihood function:
\begin{equation}
  L_{\textrm{ranking}} = -\frac{1}{B}\sum_{i=1}^{B}y_i \log{ (\hat y_i)} + (1-y_i) \log{(1-\hat y_i)},
\end{equation}
where $B$ is the batch size, $y_i \in \{0, 1\} $ is the class label denoting whether a click or conversion happens, $ \hat y_i $ is the softmaxed output of DNN. 

Due to the confidentiality policy, we can't show much details of the ranking models, so we only use a simple model to represent our ranking model. Actually, it is a complicated and large-scale ID-based model. It can be replaced to any other models in practice.

\subsection{Multimodal Model}
\label{chap:multimodal model}
\subsubsection{Modalities}
We select two modalities: visual modality from video frames, and text modality from titles, descriptions, ASR and OCR. Given the item $A$, we use $ \textrm{img}_m(A) $ and $ \textrm{txt}_k(A) $ to represent the raw content materials. Next, we extract single-modal features from vision encoder $f_{\textrm{ve}}$ and text encoder $ f_{\textrm{te}} $:
\begin{equation}
\overline{v}_{m}(A) = f_{\textrm{ve}} 
  \left( 
    \textrm{img}_m \left(
      A 
    \right) 
  \right)
, 
\quad m=1, 2, ..., M,
\end{equation}
\begin{equation}
\overline{t}_{k}(A) = f_{\textrm{te}} 
  \left( 
    \textrm{txt}_k \left(
      A 
    \right) 
  \right)
, 
\quad k=1, 2, ..., K,
\end{equation}
where $M$ indicates the number of visual modalities, $K$ indicates the number of text modalities, $ \overline{v}_m(A) $ and $ \overline{t}_k(A) $ represent the output of single-modal encoder. 

Then, we downsize their dimensions through several FC layers:
\begin{equation}
  v_{m}(A) = f_v \left(
    \overline{v}_{m} \left( A \right)
  \right),
\end{equation}
\begin{equation}
  t_{k}(A) = f_t \left(
    \overline{t}_{k} \left( A \right)
  \right),
\end{equation}
in which $ f_v $ and $ f_t $ represent the FC layers,  $ {v}_m(A) $ and $ {t}_k(A) $ are the final single-modal representations.

\subsubsection{Multimodal Fusion}
Motivated by \cite{li2023blip}, we propose Fusion-Q-Former (FQ-Former) to fuse different modalities, which is made up of transformers and a set of trainable queries. 

Given the sets of single-modalities 
$ \boldsymbol{v_A} = \left\{ v_{1}(A),  \dots, v_{M}(A) \right\} $ and 
$ \boldsymbol{t_A} = \left\{ t_{1}(A),  \dots ,t_{K}(A) \right\} $, 
we concatenate them with the globally shared queries $ \boldsymbol q= \left\{ q_1, \dots, q_{Q} \right\} $, where $Q$ is the number of queries. 
Then we input them into the transformers, which are abbreviated as $ \textrm{TRM} $ in \cref{eq: fqfomer}, aiming to learn the relationships and importance between different modalities using self-attention (SA). Finally, we slice the first $ Q $ output tokens to generate the multimodal content features $c_A$:
\vspace{-0.5em}
\begin{equation}
    \label{eq: fqfomer}
    \begin{aligned}
        c_A &= f_\textrm{FQ-Former} \left(
          \boldsymbol q, \boldsymbol v_A, \boldsymbol t_A
        \right) 
        \\
        &= \textrm{TRM} \left(
          \textrm{concat}\left(
            \left[
              \boldsymbol q, \boldsymbol v_A, \boldsymbol t_A
            \right]
          \right)
        \right) 
        \left[ : Q \right].
    \end{aligned}
\end{equation}

FQ-Former has advantages over traditional fusion methods: (i) \emph{Fixed-length}: the output size of FQ-Former is fixed and regardless of the number of modalities, making it more suitable for the industrial variable-length modalities. (ii) \emph{Robust}: since the queries participate in SA, FQ-Former can relieve the potential negative impacts, which might be caused by the low-quality materials or already enough interaction data, by assigning more attention weight to queries.

\subsubsection{Hybrid Training}
It is widely acknowledged that recent open-source backbones have been sufficiently trained and very close to the ceiling. Therefore, freezing the weights in single-modal encoders and only training the top layers or additional modules are gradually adopted by recent works \cite{ge2018image, chen2022hybrid, hu2021lora, li2023blip}. We also choose this hybrid training paradigm by freezing $ f_{\textrm{ve}} $ and $f_{\textrm{te}} $; and end-to-end training the $ f_v $, $f_t$, $\boldsymbol q $ and $ f_{\textrm{FQ-Former}} $, to balance the trade-off between efficiency and effectiveness.

\subsection{Sequential Modeling}
\label{chap:sequential modeling}

\begin{figure}[t]
  \centering
  \includegraphics[width=120px,height=120px]{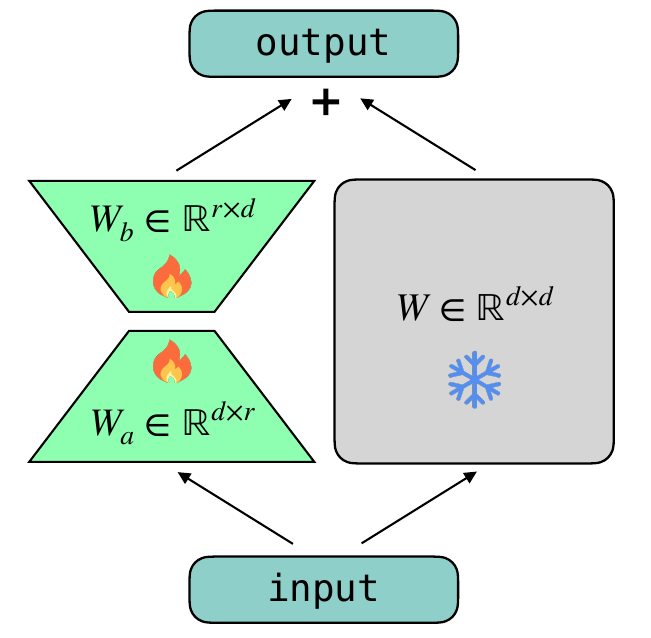}
  \caption{The details of LoRA module. The right $W ^ {d \times d}$ is frozen, while the left $W_a ^{d\times r}$ and $W_b^{r\times d}$ are trainable.
  }
  \Description{LoRA}
  \label{fig:lora}
\vspace{-0.7em}
\end{figure}

\subsubsection{User Content Interest }
Earlier works \cite{zhou2018deep, pi2020search} have shown that sequential modeling can improve the performance, guiding us to model user content interest in a similar manner. 

Given the candidate item $A$ and the user behavior sequence $\boldsymbol{u} =\{u_1, u_2, ..., u_N\}$, we first utilize the multimodal model to generate their content embeddings $c_A$ and $\{c_{u_1}, c_{u_2},...,c_{u_N} \}$. 
Then we calculate the attention score between $c_A$ and each $c_{u_{i}}$, and utilize the scores to weighted average the content sequence:
\vspace{-0.5em}
\begin{equation}
    u_A=f \left(c_A, c_{u_1}, c_{u_2},...,c_{u_N} \right)
    =\sum_{i=1}^{N}a \left(c_A, c_{u_i} \right)c_{u_i}=\sum_{i=1}^{N}w_i c_{u_i} ,
\end{equation}
\begin{equation}
    w_i = \frac{
      \exp \left(
        {a \left(c_A, c_{u_i} \right)} 
       \right)
    }
    {
      \sum_{j=1}^{N}
        \exp \left(
          {a \left(c_A, c_{u_i} \right)} 
        \right)
    },
\end{equation}
where $u_A$ represents the user content interest, a($\cdot$) represents the attention score function, $N$ is the sequence length, $w_i$ is the softmaxed attention score.

\subsubsection{LoRA-based Long-term Content Interest}
In practice, we encounter difficulties with GPU OOM when we intend to increase the sequence length. Given the hidden-size $d$, each FC layer in transformers requires $O(d^2)$ consumption. Taking the sequence length $N$ into consideration, the consumption inflates to $O(N\cdot d^2)$, greatly limiting the sequence length.

We notice that the parameters in multimodal model will stabilize after a few days of training, which inspires us to utilize the Low-Rank Adaptation (LoRA) technique \cite{hu2021lora} by switching from full tuning to partial tuning. Let's consider the weights $W ^ {d \times d}$ that need to be optimized, we first train it on a short sequence length. After a period of sufficient training, we freeze $W$ and add a trainable LoRA module to continuously follow the time-varying online distribution.  

As shown in \cref{fig:lora}, the LoRA module contains two bypassed FC layers $W_a^{d\times r}$ and $W_b^{r\times d}$, where $ r\ll d$. Compared with  the $W ^ {d \times d}$, they have the same output size but fewer trainable parameters. We add their outputs together: 
\begin{equation}
    f_{\textrm{LoRA} }(x) = \textrm{sg}(Wx)+W_b(W_ax),
\end{equation}
where $\textrm{sg} (\cdot )$ is the stop gradient operator. By the way, we reduce the number of trainable parameters from $O(d^2)$ to $O(rd)$, allowing us to increase the sequence length $N$ so that we can model user long-term content interest.

\subsection{Content-ID-Contrastive Learning}
\label{chap:cic}

As we know, ID embeddings have good memory and outperform on popular items, but have cold-start issues. Content embeddings are more generalizable, but they cannot benefit from interaction data. We propose a Content-ID-Contrastive (CIC) learning task to complement their advantages. 

Given the item $i$, we first select several important ID embeddings (e.g. ItemID, CategoryID) and concatenate them as  $\textrm{id}_i$. Next, we linearly transform the content embedding $c_i$ and ID embeddings $\textrm{id}_i$ into the same vector space:
\begin{equation}
    C_i = f_{\textrm{ CIC }}(c_i) ,
\end{equation}
\begin{equation}
    I_i = f_{\textrm{ CIC }}'(\textrm{id}_i) ,
\end{equation}
where $f_{\textrm{CIC} }$ and $f_{\textrm{CIC}}'$ are the FC layers, $C_i$ and $I_i$ are the output embeddings with the same dimension. They are positive samples of each other in the following contrastive learning.

We randomly choose $H$ negative samples for each $C_i$ and $I_i$ from the training batch, 
which are defined as $ \boldsymbol{I_{i}^{-}} =\{ I_{i1}^-, I_{i1}^-, ..., I_{iH}^- \} $  
and $ \boldsymbol{C_{i}^{-}} =\{ C_{i1}^-, C_{i1}^-, ..., C_{iH}^- \} $.
Then we utilize negative log-likelihood function to maximize the similarities of each positive pair and minimize the similarities of negative pairs:
\begin{equation}
    \begin{tiny}
    L_{\textrm{C2I} } = 
    -\frac{1}{B}\sum_{i=1}^{B}
      \log \frac{
        \exp \left(s\left(C_i, I_i \right)/\tau \right)
      }{
        \exp(s(C_i, I_i)/\tau) + \sum_{j=1}^H \exp(s(C_i, I_{ij}^-)/\tau)
    } ,
    \end{tiny}
\end {equation}
\begin{equation}
    \begin{tiny}
    L_{\textrm{I2C} } = 
    -\frac{1}{B}\sum_{i=1}^{B}
      \log \frac{
        \exp \left(s\left(I_i, C_i \right)/\tau \right)
      }{
        \exp(s(I_i, C_i)/\tau) + \sum_{j=1}^H \exp(s(I_i, C_{ij}^-)/\tau)
    } ,
    \end{tiny}
\end {equation}
where $s(\cdot)$ represents the cosine similarity, $\tau$ is the temperature parameter. Finally, we add their average to the ranking loss: 
\begin{equation}
    L = L_{\textrm{ranking} } + \alpha \cdot L_{\textrm {CIC} }= 
        L_{\textrm{ranking} } + 0.5 \alpha \cdot (L_{C2I} + L_{I2C}),
\end {equation}
where $\alpha$ is a hyperparameter. 

Considering the constraint of both ranking loss and CIC loss, CIC has different effects on different items:
\begin{itemize}[leftmargin=15px]
    \renewcommand{\labelenumi}{\roman{enumi}.}
    
    \item On \emph{popular items}: ID will dominate the alignment and inject user behavioral information into the content embeddings, generating more task-oriented content embeddings and modeling user interest better.

    \item On \emph{cold-start items}: content will dominate the alignment and guide the update of ID. As a result, the items with similar materials will also have similar ID embeddings, which promotes the generalization of RS. 
\end{itemize}


\subsection{Feature System}
\label{chap:feature system}

\begin{figure}[htb]
\vspace{-0.5em}
  \centering
  \includegraphics[width=\linewidth]{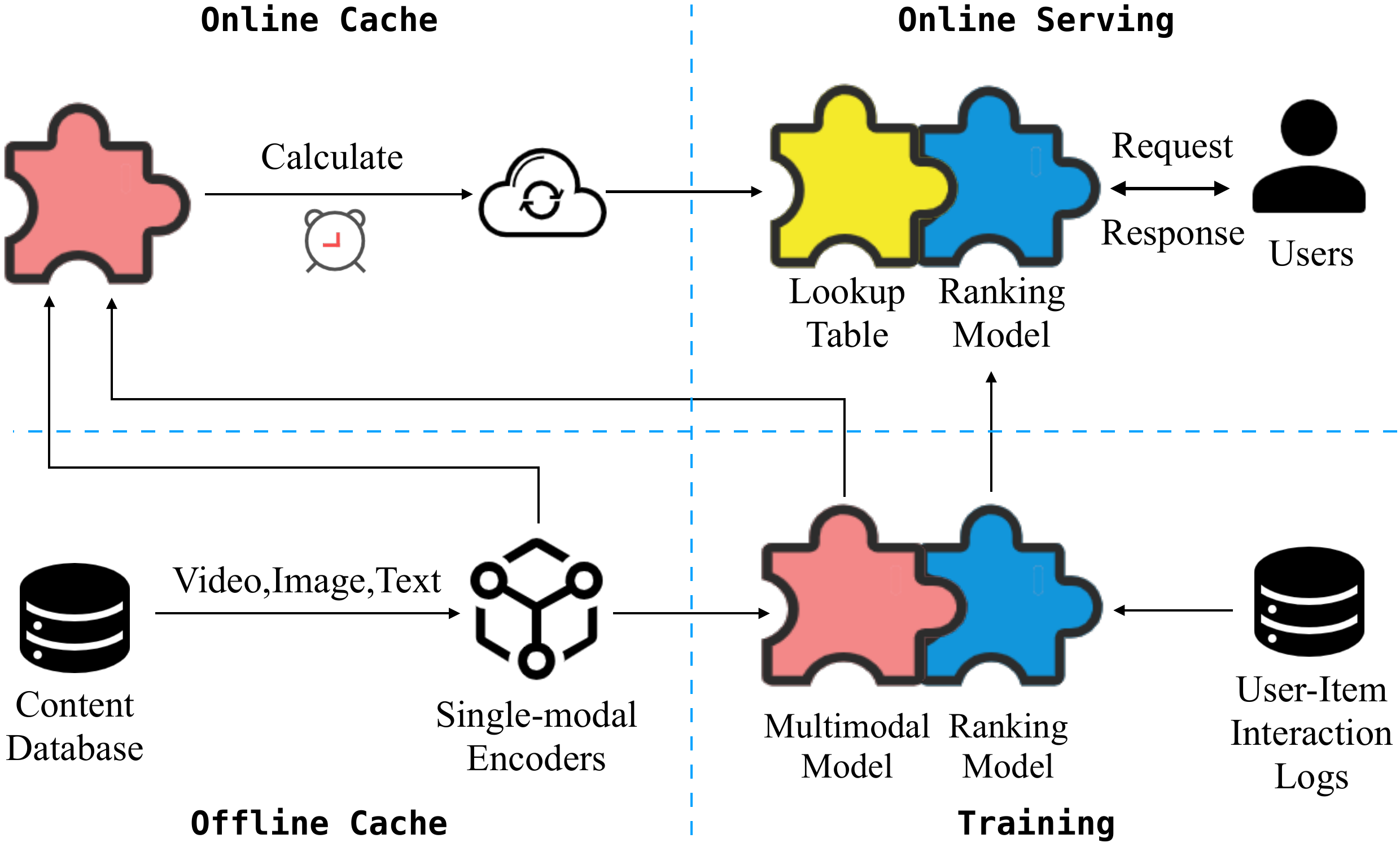}
  \caption{The feature system of EM3.}
  \Description{contains four modules: a) Offline cache: precalculates the single-modal features of every item. b) Training: trains the multimodal model with ranking model in end-to-end paradigm. c) Online cache: infers and caches fused embeddings at regular intervals. d) Online serving: based on the cached features, ranking model can get recommendation result at a very low time cost.}
  \label{fig:feature system}
\vspace{-0.5em}
\end{figure}

We design the feature system as shown in \cref{fig:feature system} to optimize the efficiency in two aspects:

\begin{itemize}[leftmargin=15px]
    \renewcommand{\labelenumi}{\roman{enumi}.}
    
    \item \emph{Training}: since the  $f_{\textrm{ve}}$ and $f_{\textrm{te}}$ are frozen, we precalculate and cache the single-modal features $\overline{v}_m(A)$ and $\overline{t}_k(A)$ to accelerate the offline training.
    
    \item \emph{Serving}: we infer and cache multimodal embeddings $ c_A $ for all items at regular intervals. When RS receiving an online request, the multimodal embeddings can be looked up directly without the forward propagation. 
\end{itemize}

\section{Experiment}
\subsection{Setup}

\subsubsection*{Baselines}
We conducted experiments on two state-of-the-art production models in our system:
\begin{itemize}[leftmargin=15px]
    \renewcommand{\labelenumi}{\roman{enumi}.}
    \item An e-com online-learning CTR model that predicts whether a user will click on the item.
    \item An ad batch-training CVR model that estimates whether a user will stay in the app in the next day.
\end{itemize}

\subsubsection*{Datasets}
Our experimental datasets come from our RS. In the e-com scenario, we use 2.1 billion records as the training set and 250 million as the test set. In the ad scenario, we use 82 million records as the training set and 2.2 million as the test set. Due to the confidentiality policy, we do not specify the beginning and ending time of dataset.

\subsubsection*{Metrics}
In offline evaluation and ablation studies, we use AUC as the primary performance indicator. In online A/B test, we consider three key metrics for the e-com system: the Gross Merchandise Value (GMV), Order Volume, and CTR. For the ad system, we consider two important metrics: the Revenue Per Mille (RPM) and Income.

\subsubsection*{Backbones}
We utilize the Swin-T-22K\footnote{\url{https://github.com/microsoft/Swin-Transformer}} as the vision encoder, and choose the RoBERTa-wwm-ext\footnote{\url{https://github.com/ymcui/Chinese-BERT-wwm}} as the text encoder. Both have been proven to be powerful single-modal models in practice.

\subsubsection*{Hyperparameters}
The values of the hyperparameters that we select are as follows: the number of queries is 2, the number of FQ-Former layers is 1, the number of SA heads is 4, and the weight of CIC loss is 0.1, the CIC temperature is 0.1.

\subsection{Offline Evaluation}
During training, the e-com model needs extra 50\% GPUs for keeping up with online-training data; the ad model does not require additional GPU. 
Both are acceptable to us.

As shown in \cref{tab:offline auc}, the e-com AUC increases by \textbf{0.256\%}, and the ad AUC increases by \textbf{0.242\%}, which are both significant under the T-test. 

\begin{table}[htb]
\vspace{-0.5em}
  \caption{Comparison of AUC.}
  \label{tab:offline auc}
  \begin{tabular}{cccc}
    \toprule
    scenario                & Method    & AUC       & AUC gain\\
    \midrule
    
    \multirow{2}{*}{e-com}  & baseline  & 0.7803    & - \\
                            & EM3       & 0.7823    & 0.256\% \\
    \hline
    \multirow{2}{*}{ad}     & baseline  & 0.7016    & - \\
                            & EM3       & 0.7033    & 0.242\% \\
  \bottomrule
\end{tabular}
\vspace{-0.8em}
\end{table}

    
     
     

\subsection{Comparison \& Ablation}
Except for the LoRA experiment, which is conducted on the e-com model, our other studies are all conducted on the ad model due to its lower experimental cost.

\subsubsection{Modalities}
We combine the following modalities in various ways to evaluate how different modalities impact the performance:
\begin{itemize}[leftmargin=15px]
    \renewcommand{\labelenumi}{\roman{enumi}.}
    \item Text: a sentence composed of title, description, ASR and OCR.
    \item Image: the cover frame of a short video.
    \item Video: several frames from a short video.
\end{itemize}

The results are listed in \cref{tab:modalities}, demonstrating that multi-modalities can facilitate the modeling on recommendation, and increasing the number of modalities can further enhance performance.
\begin{table}[htb]
\vspace{-0.5em}
  \caption{Comparison of modalities.}
  \label{tab:modalities}
  \begin{tabular}{ccc}
    \toprule
    method          & AUC       & AUC gain\\
    \midrule
    baseline        & 0.7016    & - \\
    only image	    & 0.7024	& 0.114\% \\
    only text	    & 0.7022	& 0.086\% \\
    image \& text	& 0.7027	& 0.157\% \\
    video \& text	& 0.7033	& 0.242\% \\
  \bottomrule
  \Description{(a) image modality: the cover frame of short video. 
  (b) text modality: a sentence that contains title, descriptions, ASR and OCR of video. 
  (c) video modality: several frames of short videos.}
\end{tabular}
\vspace{-1.5em}
\end{table}

\subsubsection{Fusion Methods}
We compare performance of the following fusion methods:
\begin{itemize}[leftmargin=15px]
    \renewcommand{\labelenumi}{\roman{enumi}.}
    \item 1flow \cite{kim2021vilt}: concatenates modalities and fuses them using transformers. The variable-length sequential outputs with paddings are used as the content features.
    \item 2flow \cite{lu2019vilbert}: utilizes two independent transformers with cross-attention layers to allow the modalities to interact with each other. The output of text flow is used as the content feature.
    \item Masked FQ-Former: the queries interact with modalities via Q \& K, but their V do not participate in pooling.
    \item FQ-Former: our proposed method.
\end{itemize}

As shown in \cref{tab:fusion}, FQ-Former outperforms traditional fusion methods, achieving the highest AUC gain, and taking queries into attentive pooling can further boost the improvements. 

\begin{table}[htb]
\vspace{-0.5em}
  \caption{Comparison of fusion methods.}
  \label{tab:fusion}
  \begin{tabular}{ccc}
    \toprule
    method         & AUC       & AUC gain\\
    \midrule
    baseline	   & 0.7016	   & - \\
    1flow	       & 0.7030	   & 0.199\% \\
    2flow	       & 0.7029	   & 0.190\% \\
    masked FQ-Former & 	0.7031 & 0.214\% \\
    FQ-Former	   & 0.7033	   & 0.242\% \\

  \bottomrule
  \Description{}
  
\end{tabular}
\vspace{-2em}
\end{table}

\subsubsection{LoRA}

We compare the AUC before and after using LoRA. As shown in \cref{tab:lora}, when training all parameters with a sequence length 20, the AUC gain is 0.179\%; when using LoRA with a sequence length 50, the AUC gain is 0.256\%,
which verifies that a longer sequence with fewer trainable parameters can improve the performance.

\begin{table}[htb]
\vspace{-0.5em}
  \caption{Ablation of LoRA.}
  \label{tab:lora}
  \begin{tabular}{ccc}
    \toprule
    method         & AUC       & AUC gain\\
    \midrule
    baseline	           & 0.7803	   & - \\
    length=20 w/o LoRA	   & 0.7817	   & 0.179\% \\
    length=50 w/ LoRA	   & 0.7823	   & 0.256\% \\
  \bottomrule
\end{tabular}
\vspace{-1em}
\end{table}

\subsubsection{CIC}

We conduct this ablation experiment to assess the improvement of CIC. As shown in \cref{tab:cic}: when training without CIC, the AUC gain is 0.143\%; when training with CIC, the AUC gain is 0.242\%, demonstrating that CIC can benefit the performance.

\begin{table}[htb]
\vspace{-0.5em}
  \caption{Ablation of CIC.}
  \label{tab:cic}
  \begin{tabular}{ccc}
    \toprule
    method         & AUC       & AUC gain\\
    \midrule
    baseline	   & 0.7016	   & - \\
    w/o CIC	       & 0.7026	   & 0.143\% \\
    w/ CIC	       & 0.7033	   & 0.242\% \\
  \bottomrule
\end{tabular}
\vspace{-1em}
\end{table}

\subsubsection{Item \& User}
The final content features in our method are composed of two parts: the item-side $c_A$ and the user-side $u_A$, this ablation study aims to understand the benefits of each. As shown in \cref{tab:item user}, either item feature or user feature can improve the performance, combining them together achieves the best.

\begin{table}[htb]
\vspace{-0.5em}
  \caption{Ablation of item \& user features.}
  \label{tab:item user}
  \begin{tabular}{ccc}
    \toprule
    method         & AUC       & AUC gain\\
    \midrule
    baseline	   & 0.7016	   & - \\
    only item	   & 0.7021	   & 0.071\% \\
    only user	   & 0.7024	   & 0.114\% \\
    item \& user   & 0.7033	   & 0.242\% \\
  \bottomrule
\end{tabular}
\vspace{-1em}
\end{table}

\subsubsection{Splitting the Gains}
\label{chap:split}

The gains of E2E may come from three aspects: (1) the injection of content information; (2) the guidance from recommendation task; (3) the parametric matching between the multimodal model and the ranking model. Accordingly, we set up three experimental groups to gradually split the benefits:
\begin{itemize}[leftmargin=15px]
    \renewcommand{\labelenumi}{\roman{enumi}.}
    \item PE: pre-extracts frozen features from a multimodal model, which has the similar structure with FQ-Former and has been fine-tuned in our system.
    \item Task-specific PE: we first fine-tune the content model through E2E method, and pre-extract content embeddings for all items. Finally, we feed them into the ranking model as frozen features.
    \item E2E: continuously trains the multimodal model with ranking model together.
\end{itemize}

As shown in \cref{tab:e2e}, the improvement of (1) is 0.071\%; (2) adds another 0.114\% increase; (3) provides an additional promotion of 0.057\%. Combining them all together can result in a total promotion of 0.242\%.


\begin{figure*}[htb]
  \subfigure[
{Showcases of cold-start items: the baseline ItemIDs always have little content relevance due to the lack of training. In contrast, benefited from CIC, the ItemIDs of EM3 can quickly converge near other items that have similar materials.}
  ]{
    \begin{minipage}[t]{0.48\textwidth}
      \centering
      \includegraphics[width=\textwidth]{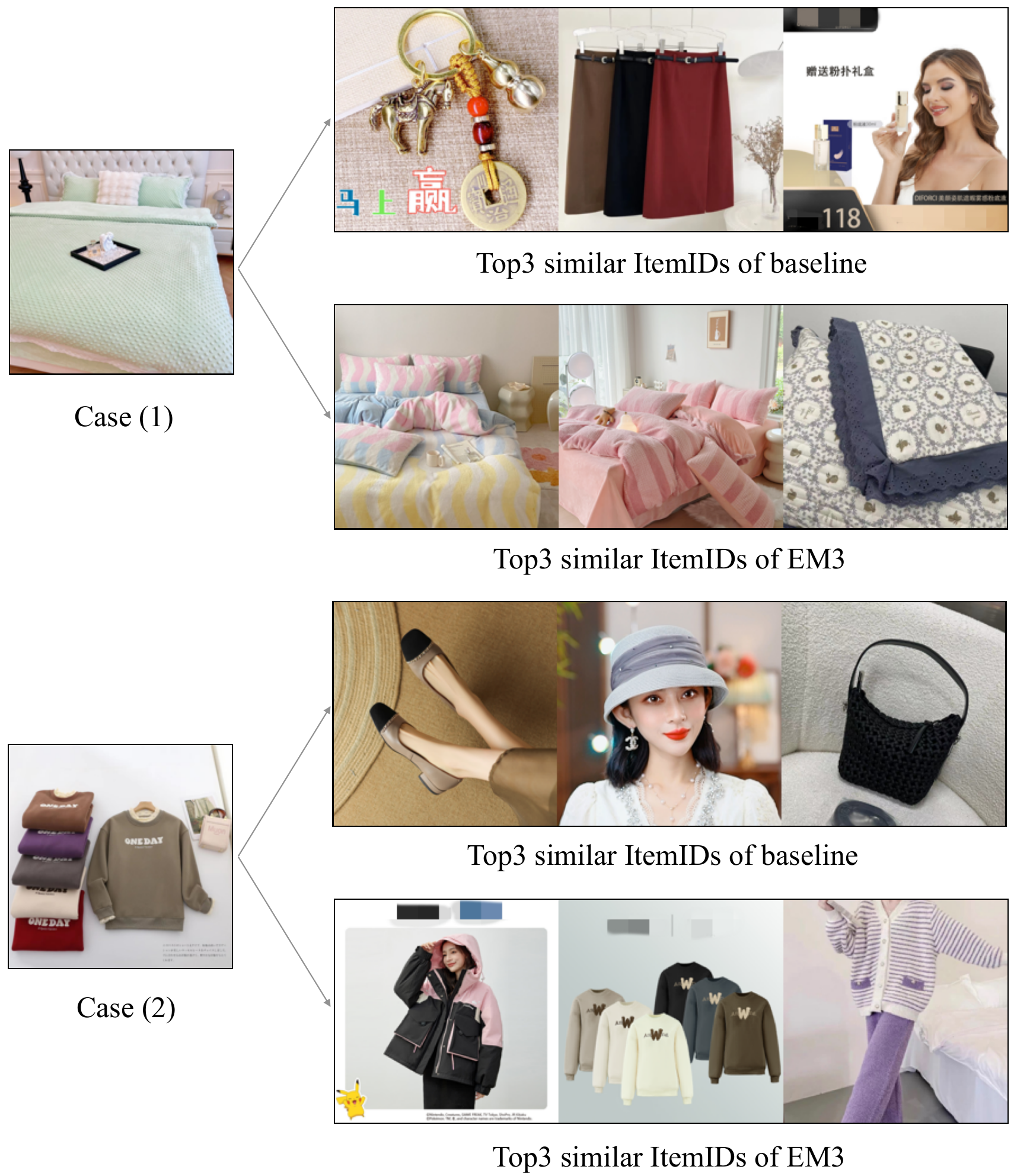}
      \label{fig:cold case}
    \end{minipage}
  }
  \hfill
  \subfigure[
  {Showcases of popular items: in case (1), EM3 captures the information that adults are the real buyers of kids' clothing; in case (2), EM3 focuses more on the products themselves rather than only on the arrangement of pictures.}
  ]{
    \begin{minipage}[t]{0.48\textwidth}
      \centering
      \includegraphics[width=\textwidth]{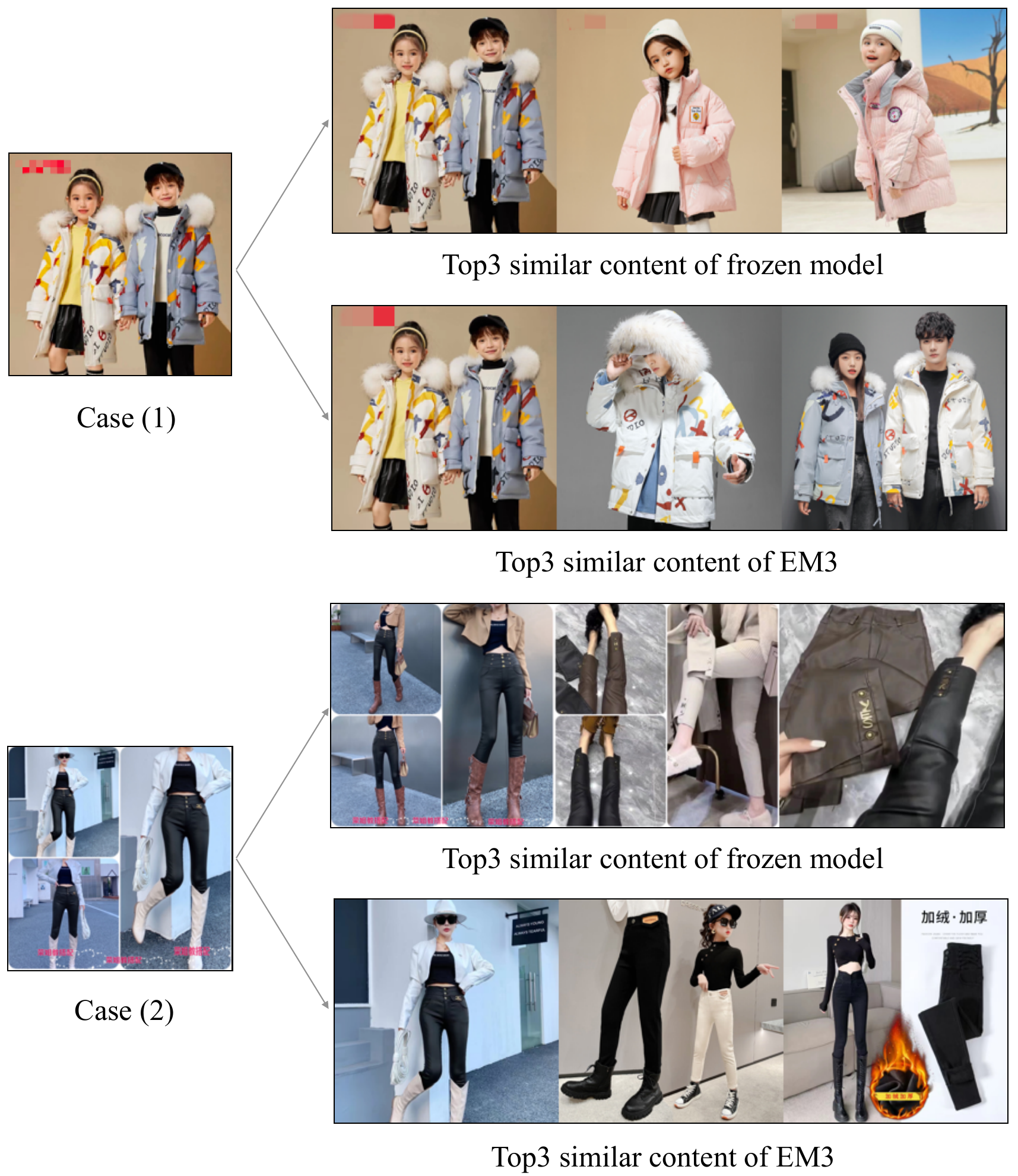}
      \label{fig:popular case}
    \end{minipage}
  }
  \caption{Visualization of the impacts on embeddings.}
  \label{fig:visualization}
\end{figure*}

\begin{table}[htb]
\vspace{-0.5em}
  \caption{Splitting the Gains of E2E.}
  \label{tab:e2e}
  \begin{tabular}{ccc}
    \toprule
    method              & AUC       & AUC gain\\
    \midrule
    baseline            & 0.7016	& - \\
    PE	                & 0.7021	& 0.071\% \\
    task-specific PE    & 0.7029	& 0.185\% \\
    E2E	                & 0.7033	& 0.242\% \\

  \bottomrule
\end{tabular}
\vspace{-0.5em}
\end{table}

\subsection{Online A/B Test}
We set 7-day online A/B Test in both scenario. In the e-com scenario, EM3 contributes to a \textbf{3.22\%} improvement on GMV, a \textbf{2.92\%} increment on order volume, and an \textbf{1.75\%} promotion on CTR. In the ad scenario, EM3 achieves a \textbf{2.64\%} improvement on RPM and generates extra \textbf{3.17\%} income. The results are all significant under the T-test.

It is worth mentioning that our method brings 2.07\% more impressions for cold-start items, which will make the platform ecology healthier in the long run.

\subsection{Impacts on Embeddings \& Visualization}
\label{chap:visualization}
This section aims to quantitatively analyze and visualize how EM3 impacts the embeddings.

\subsubsection{Cold-start Items}
Cold-start IDs always lack generalization. In order to analyze the differences in ItemID embeddings between the baseline ranking model and EM3, we dump all their ItemID embeddings from the parameter server. 
Next, we randomly sample thousands of cold-start items that are created within 5 days, and search for similar items in their respective vector spaces using cosine similarity. 
Then we calculate the material similarities between each candidate item and its top3 similar items. 
This is done in an indifferent vector space from a frozen multimodal model, which has been fine-tuned in our system and has been utilized in \cref{chap:split}. 

As the results shown in \cref{tab:cold-start items} and the typical cases shown in \cref{fig:cold case}, the cold-start ItemIDs of EM3 achieve a great improvement in material similarity.

\begin{table}[htb]
\vspace{-0.5em}
  \caption{Material similarity of ItemID.}
  \label{tab:cold-start items}
  \begin{tabular}{ccc}
    \toprule
    model               & material similarity    & gain\\
    \midrule
    baseline	& 0.3185	            & - \\
    EM3	        & 0.3363	            & 5.588\% \\
    
  \bottomrule
\end{tabular}
\vspace{-1em}
\end{table}

\subsubsection{Popular Items}
On popular items, we mainly focus on whether behavioral information has an impact on the content embeddings. 
For fairness, we choose the frozen multimodal model mentioned above as a comparison for EM3.
First, we utilize the frozen model and EM3 to calculate content embeddings for all items.
Next, we randomly sample thousands of popular items from top 30\% GMV, and search for similar items in their respective vector spaces. 
Then we calculate the behavioral similarities between each candidate item and its top3 similar items. 
This is done in the vector space of baseline ItemID embeddings, which have only been trained by user-item interaction data and can be used to measure the behavioral similarity.

As the results shown in \cref{tab:popular items} and the typical cases shown in \cref{fig:popular case}, we can conclude that the content embeddings of EM3 achieve a significant improvement in behavioral similarity.

\begin{table}[htb]
\vspace{-0.5em}
  \caption{Behavioral similarity of content embeddings.}
  \label{tab:popular items}
  \begin{tabular}{ccc}
    \toprule
    model                       & behavioral similarity & gain\\
    \midrule
    frozen	& 0.2275	            & - \\
    EM3	    & 0.2514	            & 10.505\% \\

  \bottomrule
\end{tabular}
\vspace{-1em}
\end{table}

\subsection{Public Evaluation}

\begin{table*}[htb]
  \caption{Performance of different recommendation models on public datasets. The best results are marked in boldface and the second best results are underlined.}
  \label{tab:public eval}
  \begin{tabular}{cccccccccc}
    \toprule
    \multirow{2}{*}{Datasets}  & \multirow{2}{*}{Metrics} & \multicolumn{2}{c}{General Models} & \multicolumn{6}{c}{Multimodal Models}\\

    \cmidrule(r){3-4}  \cmidrule(r){5-10}

            & & BPR	    & LightGCN      & VBPR      & SLMRec    & LATTICE   & BM3   & FREEDOM & EM3 \\

    \midrule
    \multirow{4}{*}{Baby} 
    & R@10 &    0.0357	 & 0.0479	  & 0.0423	    & 0.0521	& 0.0551	& 0.0564	& \underline{0.0624}   & \textbf{0.0646} \\
    & R@20 &    0.0575	 & 0.0754	  & 0.0663	    & 0.0772    & 0.0852	& 0.0883	& \underline{0.0980}   & \textbf{0.1032} \\
    & N@10 &    0.0192	 & 0.0257	  & 0.0223	    & 0.0289	& 0.0292	& 0.0301	& \underline{0.0324}   & \textbf{0.0336} \\
    & N@20 &    0.0249	 & 0.0328	  & 0.0285	    & 0.0354	& 0.0369	& 0.0383	& \underline{0.0416}   & \textbf{0.0435} \\

    \midrule

    \multirow{4}{*}{Sports} 
    & R@10 &   0.0432    & 0.0569	  & 0.0560	    & 0.0663	& 0.0621	& 0.0656	& \underline{0.0713}   & \textbf{0.0726} \\
    & R@20 &   0.0653	 & 0.0864	  & 0.0854	    & 0.0990	& 0.0957	& 0.0980	& \underline{0.1075}   & \textbf{0.1099} \\
    & N@10 &   0.0241	 & 0.0311	  & 0.0307	    & 0.0365	& 0.0335	& 0.0355	& \underline{0.0384}   & \textbf{0.0391} \\
    & N@20 &   0.0298	 & 0.0387	  & 0.0383	    & 0.0450	& 0.0422	& 0.0438	& \underline{0.0477}   & \textbf{0.0488} \\

  \bottomrule
\end{tabular}
\end{table*}

Because our proposed modules can be also combined with some two-tower models, we conduct experiments on two public datasets. Thanks to the MMRec\footnote{\url{https://github.com/enoche/MMRec}} framework contributed by \cite{zhou2023comprehensive}, we can test at a very low cost.

\subsubsection*{Datasets}
The Amazon dataset has been widely used in previous studies, providing both interaction records and multimodal information. We choose 2 categories: Baby and Sports. For each category, we randomly split 80\% of historical records as a training set, 10\% for validation and the remaining 10\% for test.

\subsubsection*{Method}
We integrate the fusion and CIC modules into the FREEDOM \cite{zhou2023tale}, which is one of the most effective works at the time of writing. In details, we first use FQ-Former to calculate the multimodal embeddings of all items. Then we utilize GCN to process them on both item-side and user-side. Next, we align the item-side multimodal embeddings with ItemID embeddings using CIC. Finally, we concatenate them for retrieving.

Same as the previous works, we fix the embedding size of both users and items to 64, and use the negative sampling strategy to pair each user-item interaction with one negative item. 

\subsubsection*{Baselines}
We compare our proposed model with the following methods: BPR \cite{rendle2012bpr}, LightGCN \cite{he2020lightgcn}, VBPR \cite{he2016vbpr}, SLMRec \cite{tao2022self}, LATTICE \cite{zhang2021mining}, BM3 \cite{zhou2023bootstrap}, FREEDOM. For a fair comparison, we retrain FREEDOM and some other models in our environment.

\subsubsection*{Hyperparameters}
We perform a grid search to find its optimal settings on different datasets: 
we search for 
the number of queries from \{1, 2\}, 
the dropout rate of transformer from \{0.5, 0.8\}, 
and the CIC temperature from \{0.5, 0.1\}.


\subsubsection*{Metrics}
We select two widely-used evaluation metrics for top-K recommendation: Recall@K and NDCG@K, which are abbreviated as R@K and N@K. We use R@20 on the validation data as the training stopping indicator, and report the average metrics of all users in the test sets for both K=10 and K=20. 
 
\subsubsection*{Results}
As shown in \cref{tab:public eval}, EM3 outperforms the baselines on both datasets. It shows that the method proposed by us is generalizable. The source code and evaluation logs are available at \url{https://github.com/em3e2e-anonymous/em3}.

Because the public evaluation is not the main focus of our research, we only simply integrate our method with one of previous works. 
Considering the flexibility of EM3 in decoupling from the original recommendation models, we believe that it can be combined with other models for a similar improvement.

\section{Conclusion}
In this paper, we propose an industrial multimodal recommendation framework named EM3 for end-to-end training of multimodal model and ranking model. EM3 sufficiently utilizes multimodalities and allows personalized ranking tasks to directly train the core modules in the multimodal model, obtaining more task-oriented content representations. In details, we propose FQ-Former to fuse different modalities and generate fixed-length and robust content embeddings. In user sequential modeling, we utilize LoRA technique to reduce the consumption of trainable parameters so that we can increase the length of behavioral sequence to model user content interest better. Besides, we propose a novel CIC learning task to complement the advantages of content and ID through alignment, which allows us to obtain more task-oriented content embeddings and more generalized ID embeddings. The experiments conducted in two different scenarios show that EM3 is achieves significant improvements and bring in millions of revenue, which also verify the generalizability of our method. The evaluation on public datasets also show that our proposed method outperforms the state-of-the-art methods.

In the future, we are going to focus on the direction of E2E training, to add more modalities such as audio, or to end-to-end train MLLM.

\bibliographystyle{ACM-Reference-Format}
\bibliography{ref}


\begin{thebibliography}{58}


\ifx \showCODEN    \undefined \def \showCODEN     #1{\unskip}     \fi
\ifx \showDOI      \undefined \def \showDOI       #1{#1}\fi
\ifx \showISBNx    \undefined \def \showISBNx     #1{\unskip}     \fi
\ifx \showISBNxiii \undefined \def \showISBNxiii  #1{\unskip}     \fi
\ifx \showISSN     \undefined \def \showISSN      #1{\unskip}     \fi
\ifx \showLCCN     \undefined \def \showLCCN      #1{\unskip}     \fi
\ifx \shownote     \undefined \def \shownote      #1{#1}          \fi
\ifx \showarticletitle \undefined \def \showarticletitle #1{#1}   \fi
\ifx \showURL      \undefined \def \showURL       {\relax}        \fi
\providecommand\bibfield[2]{#2}
\providecommand\bibinfo[2]{#2}
\providecommand\natexlab[1]{#1}
\providecommand\showeprint[2][]{arXiv:#2}

\bibitem[Bahdanau et~al\mbox{.}(2014)]%
        {bahdanau2014neural}
\bibfield{author}{\bibinfo{person}{Dzmitry Bahdanau}, \bibinfo{person}{Kyunghyun Cho}, {and} \bibinfo{person}{Yoshua Bengio}.} \bibinfo{year}{2014}\natexlab{}.
\newblock \showarticletitle{Neural machine translation by jointly learning to align and translate}.
\newblock \bibinfo{journal}{\emph{arXiv preprint arXiv:1409.0473}} (\bibinfo{year}{2014}).
\newblock


\bibitem[Chen et~al\mbox{.}(2021)]%
        {chen2021efficient}
\bibfield{author}{\bibinfo{person}{Jin Chen}, \bibinfo{person}{Tiezheng Ge}, \bibinfo{person}{Gangwei Jiang}, \bibinfo{person}{Zhiqiang Zhang}, \bibinfo{person}{Defu Lian}, {and} \bibinfo{person}{Kai Zheng}.} \bibinfo{year}{2021}\natexlab{}.
\newblock \showarticletitle{Efficient Optimal Selection for Composited Advertising Creatives with Tree Structure}. In \bibinfo{booktitle}{\emph{Proceedings of the AAAI Conference on Artificial Intelligence}}, Vol.~\bibinfo{volume}{35}. \bibinfo{pages}{3967--3975}.
\newblock


\bibitem[Chen et~al\mbox{.}(2016)]%
        {chen2016deep}
\bibfield{author}{\bibinfo{person}{Junxuan Chen}, \bibinfo{person}{Baigui Sun}, \bibinfo{person}{Hao Li}, \bibinfo{person}{Hongtao Lu}, {and} \bibinfo{person}{Xian-Sheng Hua}.} \bibinfo{year}{2016}\natexlab{}.
\newblock \showarticletitle{Deep ctr prediction in display advertising}. In \bibinfo{booktitle}{\emph{Proceedings of the 24th ACM international conference on Multimedia}}. \bibinfo{pages}{811--820}.
\newblock


\bibitem[Chen et~al\mbox{.}(2022)]%
        {chen2022hybrid}
\bibfield{author}{\bibinfo{person}{Xin Chen}, \bibinfo{person}{Qingtao Tang}, \bibinfo{person}{Ke Hu}, \bibinfo{person}{Yue Xu}, \bibinfo{person}{Shihang Qiu}, \bibinfo{person}{Jia Cheng}, {and} \bibinfo{person}{Jun Lei}.} \bibinfo{year}{2022}\natexlab{}.
\newblock \showarticletitle{Hybrid CNN Based Attention with Category Prior for User Image Behavior Modeling}. In \bibinfo{booktitle}{\emph{Proceedings of the 45th International ACM SIGIR Conference on Research and Development in Information Retrieval}}. \bibinfo{pages}{2336--2340}.
\newblock


\bibitem[Cheng et~al\mbox{.}(2016)]%
        {cheng2016wide}
\bibfield{author}{\bibinfo{person}{Heng-Tze Cheng}, \bibinfo{person}{Levent Koc}, \bibinfo{person}{Jeremiah Harmsen}, \bibinfo{person}{Tal Shaked}, \bibinfo{person}{Tushar Chandra}, \bibinfo{person}{Hrishi Aradhye}, \bibinfo{person}{Glen Anderson}, \bibinfo{person}{Greg Corrado}, \bibinfo{person}{Wei Chai}, \bibinfo{person}{Mustafa Ispir}, {et~al\mbox{.}}} \bibinfo{year}{2016}\natexlab{}.
\newblock \showarticletitle{Wide \& deep learning for recommender systems}. In \bibinfo{booktitle}{\emph{Proceedings of the 1st workshop on deep learning for recommender systems}}. \bibinfo{pages}{7--10}.
\newblock


\bibitem[Cui et~al\mbox{.}(2022)]%
        {cui2022m6}
\bibfield{author}{\bibinfo{person}{Zeyu Cui}, \bibinfo{person}{Jianxin Ma}, \bibinfo{person}{Chang Zhou}, \bibinfo{person}{Jingren Zhou}, {and} \bibinfo{person}{Hongxia Yang}.} \bibinfo{year}{2022}\natexlab{}.
\newblock \showarticletitle{M6-rec: Generative pretrained language models are open-ended recommender systems}.
\newblock \bibinfo{journal}{\emph{arXiv preprint arXiv:2205.08084}} (\bibinfo{year}{2022}).
\newblock


\bibitem[Devlin et~al\mbox{.}(2018)]%
        {devlin2018bert}
\bibfield{author}{\bibinfo{person}{Jacob Devlin}, \bibinfo{person}{Ming-Wei Chang}, \bibinfo{person}{Kenton Lee}, {and} \bibinfo{person}{Kristina Toutanova}.} \bibinfo{year}{2018}\natexlab{}.
\newblock \showarticletitle{Bert: Pre-training of deep bidirectional transformers for language understanding}.
\newblock \bibinfo{journal}{\emph{arXiv preprint arXiv:1810.04805}} (\bibinfo{year}{2018}).
\newblock


\bibitem[Dosovitskiy et~al\mbox{.}(2020)]%
        {dosovitskiy2020image}
\bibfield{author}{\bibinfo{person}{Alexey Dosovitskiy}, \bibinfo{person}{Lucas Beyer}, \bibinfo{person}{Alexander Kolesnikov}, \bibinfo{person}{Dirk Weissenborn}, \bibinfo{person}{Xiaohua Zhai}, \bibinfo{person}{Thomas Unterthiner}, \bibinfo{person}{Mostafa Dehghani}, \bibinfo{person}{Matthias Minderer}, \bibinfo{person}{Georg Heigold}, \bibinfo{person}{Sylvain Gelly}, {et~al\mbox{.}}} \bibinfo{year}{2020}\natexlab{}.
\newblock \showarticletitle{An image is worth 16x16 words: Transformers for image recognition at scale}.
\newblock \bibinfo{journal}{\emph{arXiv preprint arXiv:2010.11929}} (\bibinfo{year}{2020}).
\newblock


\bibitem[Elsayed et~al\mbox{.}(2022)]%
        {elsayed2022end}
\bibfield{author}{\bibinfo{person}{Shereen Elsayed}, \bibinfo{person}{Lukas Brinkmeyer}, {and} \bibinfo{person}{Lars Schmidt-Thieme}.} \bibinfo{year}{2022}\natexlab{}.
\newblock \showarticletitle{End-to-end image-based fashion recommendation}. In \bibinfo{booktitle}{\emph{Workshop on Recommender Systems in Fashion and Retail}}. Springer, \bibinfo{pages}{109--119}.
\newblock


\bibitem[Ge et~al\mbox{.}(2018)]%
        {ge2018image}
\bibfield{author}{\bibinfo{person}{Tiezheng Ge}, \bibinfo{person}{Liqin Zhao}, \bibinfo{person}{Guorui Zhou}, \bibinfo{person}{Keyu Chen}, \bibinfo{person}{Shuying Liu}, \bibinfo{person}{Huimin Yi}, \bibinfo{person}{Zelin Hu}, \bibinfo{person}{Bochao Liu}, \bibinfo{person}{Peng Sun}, \bibinfo{person}{Haoyu Liu}, {et~al\mbox{.}}} \bibinfo{year}{2018}\natexlab{}.
\newblock \showarticletitle{Image matters: Visually modeling user behaviors using advanced model server}. In \bibinfo{booktitle}{\emph{Proceedings of the 27th ACM International Conference on Information and Knowledge Management}}. \bibinfo{pages}{2087--2095}.
\newblock


\bibitem[Geng et~al\mbox{.}(2022)]%
        {geng2022recommendation}
\bibfield{author}{\bibinfo{person}{Shijie Geng}, \bibinfo{person}{Shuchang Liu}, \bibinfo{person}{Zuohui Fu}, \bibinfo{person}{Yingqiang Ge}, {and} \bibinfo{person}{Yongfeng Zhang}.} \bibinfo{year}{2022}\natexlab{}.
\newblock \showarticletitle{Recommendation as language processing (rlp): A unified pretrain, personalized prompt \& predict paradigm (p5)}. In \bibinfo{booktitle}{\emph{Proceedings of the 16th ACM Conference on Recommender Systems}}. \bibinfo{pages}{299--315}.
\newblock


\bibitem[Guo et~al\mbox{.}(2017)]%
        {guo2017deepfm}
\bibfield{author}{\bibinfo{person}{Huifeng Guo}, \bibinfo{person}{Ruiming Tang}, \bibinfo{person}{Yunming Ye}, \bibinfo{person}{Zhenguo Li}, {and} \bibinfo{person}{Xiuqiang He}.} \bibinfo{year}{2017}\natexlab{}.
\newblock \showarticletitle{DeepFM: a factorization-machine based neural network for CTR prediction}.
\newblock \bibinfo{journal}{\emph{arXiv preprint arXiv:1703.04247}} (\bibinfo{year}{2017}).
\newblock


\bibitem[He et~al\mbox{.}(2016)]%
        {he2016sherlock}
\bibfield{author}{\bibinfo{person}{Ruining He}, \bibinfo{person}{Chunbin Lin}, \bibinfo{person}{Jianguo Wang}, {and} \bibinfo{person}{Julian McAuley}.} \bibinfo{year}{2016}\natexlab{}.
\newblock \showarticletitle{Sherlock: sparse hierarchical embeddings for visually-aware one-class collaborative filtering}.
\newblock \bibinfo{journal}{\emph{arXiv preprint arXiv:1604.05813}} (\bibinfo{year}{2016}).
\newblock


\bibitem[He and McAuley(2016)]%
        {he2016vbpr}
\bibfield{author}{\bibinfo{person}{Ruining He} {and} \bibinfo{person}{Julian McAuley}.} \bibinfo{year}{2016}\natexlab{}.
\newblock \showarticletitle{VBPR: visual bayesian personalized ranking from implicit feedback}. In \bibinfo{booktitle}{\emph{Proceedings of the AAAI conference on artificial intelligence}}, Vol.~\bibinfo{volume}{30}.
\newblock


\bibitem[He et~al\mbox{.}(2020)]%
        {he2020lightgcn}
\bibfield{author}{\bibinfo{person}{Xiangnan He}, \bibinfo{person}{Kuan Deng}, \bibinfo{person}{Xiang Wang}, \bibinfo{person}{Yan Li}, \bibinfo{person}{Yongdong Zhang}, {and} \bibinfo{person}{Meng Wang}.} \bibinfo{year}{2020}\natexlab{}.
\newblock \showarticletitle{Lightgcn: Simplifying and powering graph convolution network for recommendation}. In \bibinfo{booktitle}{\emph{Proceedings of the 43rd International ACM SIGIR conference on research and development in Information Retrieval}}. \bibinfo{pages}{639--648}.
\newblock


\bibitem[Hu et~al\mbox{.}(2021)]%
        {hu2021lora}
\bibfield{author}{\bibinfo{person}{Edward~J Hu}, \bibinfo{person}{Yelong Shen}, \bibinfo{person}{Phillip Wallis}, \bibinfo{person}{Zeyuan Allen-Zhu}, \bibinfo{person}{Yuanzhi Li}, \bibinfo{person}{Shean Wang}, \bibinfo{person}{Lu Wang}, {and} \bibinfo{person}{Weizhu Chen}.} \bibinfo{year}{2021}\natexlab{}.
\newblock \showarticletitle{Lora: Low-rank adaptation of large language models}.
\newblock \bibinfo{journal}{\emph{arXiv preprint arXiv:2106.09685}} (\bibinfo{year}{2021}).
\newblock


\bibitem[Jabeen et~al\mbox{.}(2023)]%
        {jabeen2023review}
\bibfield{author}{\bibinfo{person}{Summaira Jabeen}, \bibinfo{person}{Xi Li}, \bibinfo{person}{Muhammad~Shoib Amin}, \bibinfo{person}{Omar Bourahla}, \bibinfo{person}{Songyuan Li}, {and} \bibinfo{person}{Abdul Jabbar}.} \bibinfo{year}{2023}\natexlab{}.
\newblock \showarticletitle{A review on methods and applications in multimodal deep learning}.
\newblock \bibinfo{journal}{\emph{ACM Transactions on Multimedia Computing, Communications and Applications}} \bibinfo{volume}{19}, \bibinfo{number}{2s} (\bibinfo{year}{2023}), \bibinfo{pages}{1--41}.
\newblock


\bibitem[Juan et~al\mbox{.}(2016)]%
        {juan2016field}
\bibfield{author}{\bibinfo{person}{Yuchin Juan}, \bibinfo{person}{Yong Zhuang}, \bibinfo{person}{Wei-Sheng Chin}, {and} \bibinfo{person}{Chih-Jen Lin}.} \bibinfo{year}{2016}\natexlab{}.
\newblock \showarticletitle{Field-aware factorization machines for CTR prediction}. In \bibinfo{booktitle}{\emph{Proceedings of the 10th ACM conference on recommender systems}}. \bibinfo{pages}{43--50}.
\newblock


\bibitem[Kim et~al\mbox{.}(2021)]%
        {kim2021vilt}
\bibfield{author}{\bibinfo{person}{Wonjae Kim}, \bibinfo{person}{Bokyung Son}, {and} \bibinfo{person}{Ildoo Kim}.} \bibinfo{year}{2021}\natexlab{}.
\newblock \showarticletitle{Vilt: Vision-and-language transformer without convolution or region supervision}. In \bibinfo{booktitle}{\emph{International Conference on Machine Learning}}. PMLR, \bibinfo{pages}{5583--5594}.
\newblock


\bibitem[Koren et~al\mbox{.}(2009)]%
        {koren2009matrix}
\bibfield{author}{\bibinfo{person}{Yehuda Koren}, \bibinfo{person}{Robert Bell}, {and} \bibinfo{person}{Chris Volinsky}.} \bibinfo{year}{2009}\natexlab{}.
\newblock \showarticletitle{Matrix factorization techniques for recommender systems}.
\newblock \bibinfo{journal}{\emph{Computer}} \bibinfo{volume}{42}, \bibinfo{number}{8} (\bibinfo{year}{2009}), \bibinfo{pages}{30--37}.
\newblock


\bibitem[Krizhevsky et~al\mbox{.}(2012)]%
        {krizhevsky2012imagenet}
\bibfield{author}{\bibinfo{person}{Alex Krizhevsky}, \bibinfo{person}{Ilya Sutskever}, {and} \bibinfo{person}{Geoffrey~E Hinton}.} \bibinfo{year}{2012}\natexlab{}.
\newblock \showarticletitle{Imagenet classification with deep convolutional neural networks}.
\newblock \bibinfo{journal}{\emph{Advances in neural information processing systems}}  \bibinfo{volume}{25} (\bibinfo{year}{2012}).
\newblock


\bibitem[Li et~al\mbox{.}(2023)]%
        {li2023blip}
\bibfield{author}{\bibinfo{person}{Junnan Li}, \bibinfo{person}{Dongxu Li}, \bibinfo{person}{Silvio Savarese}, {and} \bibinfo{person}{Steven Hoi}.} \bibinfo{year}{2023}\natexlab{}.
\newblock \showarticletitle{Blip-2: Bootstrapping language-image pre-training with frozen image encoders and large language models}.
\newblock \bibinfo{journal}{\emph{arXiv preprint arXiv:2301.12597}} (\bibinfo{year}{2023}).
\newblock


\bibitem[Li et~al\mbox{.}(2022)]%
        {li2022blip}
\bibfield{author}{\bibinfo{person}{Junnan Li}, \bibinfo{person}{Dongxu Li}, \bibinfo{person}{Caiming Xiong}, {and} \bibinfo{person}{Steven Hoi}.} \bibinfo{year}{2022}\natexlab{}.
\newblock \showarticletitle{Blip: Bootstrapping language-image pre-training for unified vision-language understanding and generation}. In \bibinfo{booktitle}{\emph{International Conference on Machine Learning}}. PMLR, \bibinfo{pages}{12888--12900}.
\newblock


\bibitem[Li et~al\mbox{.}(2021)]%
        {li2021align}
\bibfield{author}{\bibinfo{person}{Junnan Li}, \bibinfo{person}{Ramprasaath Selvaraju}, \bibinfo{person}{Akhilesh Gotmare}, \bibinfo{person}{Shafiq Joty}, \bibinfo{person}{Caiming Xiong}, {and} \bibinfo{person}{Steven Chu~Hong Hoi}.} \bibinfo{year}{2021}\natexlab{}.
\newblock \showarticletitle{Align before fuse: Vision and language representation learning with momentum distillation}.
\newblock \bibinfo{journal}{\emph{Advances in neural information processing systems}}  \bibinfo{volume}{34} (\bibinfo{year}{2021}), \bibinfo{pages}{9694--9705}.
\newblock


\bibitem[Liu et~al\mbox{.}(2020)]%
        {liu2020category}
\bibfield{author}{\bibinfo{person}{Hu Liu}, \bibinfo{person}{Jing Lu}, \bibinfo{person}{Hao Yang}, \bibinfo{person}{Xiwei Zhao}, \bibinfo{person}{Sulong Xu}, \bibinfo{person}{Hao Peng}, \bibinfo{person}{Zehua Zhang}, \bibinfo{person}{Wenjie Niu}, \bibinfo{person}{Xiaokun Zhu}, \bibinfo{person}{Yongjun Bao}, {et~al\mbox{.}}} \bibinfo{year}{2020}\natexlab{}.
\newblock \showarticletitle{Category-Specific CNN for Visual-aware CTR Prediction at JD. com}. In \bibinfo{booktitle}{\emph{Proceedings of the 26th ACM SIGKDD international conference on knowledge discovery \& data mining}}. \bibinfo{pages}{2686--2696}.
\newblock


\bibitem[Liu et~al\mbox{.}(2023)]%
        {liu2023megcf}
\bibfield{author}{\bibinfo{person}{Kang Liu}, \bibinfo{person}{Feng Xue}, \bibinfo{person}{Dan Guo}, \bibinfo{person}{Le Wu}, \bibinfo{person}{Shujie Li}, {and} \bibinfo{person}{Richang Hong}.} \bibinfo{year}{2023}\natexlab{}.
\newblock \showarticletitle{Megcf: Multimodal entity graph collaborative filtering for personalized recommendation}.
\newblock \bibinfo{journal}{\emph{ACM Transactions on Information Systems}} \bibinfo{volume}{41}, \bibinfo{number}{2} (\bibinfo{year}{2023}), \bibinfo{pages}{1--27}.
\newblock


\bibitem[Liu et~al\mbox{.}(2019)]%
        {liu2019roberta}
\bibfield{author}{\bibinfo{person}{Yinhan Liu}, \bibinfo{person}{Myle Ott}, \bibinfo{person}{Naman Goyal}, \bibinfo{person}{Jingfei Du}, \bibinfo{person}{Mandar Joshi}, \bibinfo{person}{Danqi Chen}, \bibinfo{person}{Omer Levy}, \bibinfo{person}{Mike Lewis}, \bibinfo{person}{Luke Zettlemoyer}, {and} \bibinfo{person}{Veselin Stoyanov}.} \bibinfo{year}{2019}\natexlab{}.
\newblock \showarticletitle{Roberta: A robustly optimized bert pretraining approach}.
\newblock \bibinfo{journal}{\emph{arXiv preprint arXiv:1907.11692}} (\bibinfo{year}{2019}).
\newblock


\bibitem[Liu et~al\mbox{.}(2021)]%
        {liu2021swin}
\bibfield{author}{\bibinfo{person}{Ze Liu}, \bibinfo{person}{Yutong Lin}, \bibinfo{person}{Yue Cao}, \bibinfo{person}{Han Hu}, \bibinfo{person}{Yixuan Wei}, \bibinfo{person}{Zheng Zhang}, \bibinfo{person}{Stephen Lin}, {and} \bibinfo{person}{Baining Guo}.} \bibinfo{year}{2021}\natexlab{}.
\newblock \showarticletitle{Swin transformer: Hierarchical vision transformer using shifted windows}. In \bibinfo{booktitle}{\emph{Proceedings of the IEEE/CVF international conference on computer vision}}. \bibinfo{pages}{10012--10022}.
\newblock


\bibitem[Lu et~al\mbox{.}(2019)]%
        {lu2019vilbert}
\bibfield{author}{\bibinfo{person}{Jiasen Lu}, \bibinfo{person}{Dhruv Batra}, \bibinfo{person}{Devi Parikh}, {and} \bibinfo{person}{Stefan Lee}.} \bibinfo{year}{2019}\natexlab{}.
\newblock \showarticletitle{Vilbert: Pretraining task-agnostic visiolinguistic representations for vision-and-language tasks}.
\newblock \bibinfo{journal}{\emph{Advances in neural information processing systems}}  \bibinfo{volume}{32} (\bibinfo{year}{2019}).
\newblock


\bibitem[McMahan et~al\mbox{.}(2013)]%
        {mcmahan2013ad}
\bibfield{author}{\bibinfo{person}{H~Brendan McMahan}, \bibinfo{person}{Gary Holt}, \bibinfo{person}{David Sculley}, \bibinfo{person}{Michael Young}, \bibinfo{person}{Dietmar Ebner}, \bibinfo{person}{Julian Grady}, \bibinfo{person}{Lan Nie}, \bibinfo{person}{Todd Phillips}, \bibinfo{person}{Eugene Davydov}, \bibinfo{person}{Daniel Golovin}, {et~al\mbox{.}}} \bibinfo{year}{2013}\natexlab{}.
\newblock \showarticletitle{Ad click prediction: a view from the trenches}. In \bibinfo{booktitle}{\emph{Proceedings of the 19th ACM SIGKDD international conference on Knowledge discovery and data mining}}. \bibinfo{pages}{1222--1230}.
\newblock


\bibitem[Mo et~al\mbox{.}(2015)]%
        {mo2015image}
\bibfield{author}{\bibinfo{person}{Kaixiang Mo}, \bibinfo{person}{Bo Liu}, \bibinfo{person}{Lei Xiao}, \bibinfo{person}{Yong Li}, {and} \bibinfo{person}{Jie Jiang}.} \bibinfo{year}{2015}\natexlab{}.
\newblock \showarticletitle{Image feature learning for cold start problem in display advertising}. In \bibinfo{booktitle}{\emph{Twenty-Fourth International Joint Conference on Artificial Intelligence}}.
\newblock


\bibitem[Ouyang et~al\mbox{.}(2022)]%
        {ouyang2022training}
\bibfield{author}{\bibinfo{person}{Long Ouyang}, \bibinfo{person}{Jeffrey Wu}, \bibinfo{person}{Xu Jiang}, \bibinfo{person}{Diogo Almeida}, \bibinfo{person}{Carroll Wainwright}, \bibinfo{person}{Pamela Mishkin}, \bibinfo{person}{Chong Zhang}, \bibinfo{person}{Sandhini Agarwal}, \bibinfo{person}{Katarina Slama}, \bibinfo{person}{Alex Ray}, {et~al\mbox{.}}} \bibinfo{year}{2022}\natexlab{}.
\newblock \showarticletitle{Training language models to follow instructions with human feedback}.
\newblock \bibinfo{journal}{\emph{Advances in Neural Information Processing Systems}}  \bibinfo{volume}{35} (\bibinfo{year}{2022}), \bibinfo{pages}{27730--27744}.
\newblock


\bibitem[Pi et~al\mbox{.}(2020)]%
        {pi2020search}
\bibfield{author}{\bibinfo{person}{Qi Pi}, \bibinfo{person}{Guorui Zhou}, \bibinfo{person}{Yujing Zhang}, \bibinfo{person}{Zhe Wang}, \bibinfo{person}{Lejian Ren}, \bibinfo{person}{Ying Fan}, \bibinfo{person}{Xiaoqiang Zhu}, {and} \bibinfo{person}{Kun Gai}.} \bibinfo{year}{2020}\natexlab{}.
\newblock \showarticletitle{Search-based user interest modeling with lifelong sequential behavior data for click-through rate prediction}. In \bibinfo{booktitle}{\emph{Proceedings of the 29th ACM International Conference on Information \& Knowledge Management}}. \bibinfo{pages}{2685--2692}.
\newblock


\bibitem[Radford et~al\mbox{.}(2021)]%
        {radford2021learning}
\bibfield{author}{\bibinfo{person}{Alec Radford}, \bibinfo{person}{Jong~Wook Kim}, \bibinfo{person}{Chris Hallacy}, \bibinfo{person}{Aditya Ramesh}, \bibinfo{person}{Gabriel Goh}, \bibinfo{person}{Sandhini Agarwal}, \bibinfo{person}{Girish Sastry}, \bibinfo{person}{Amanda Askell}, \bibinfo{person}{Pamela Mishkin}, \bibinfo{person}{Jack Clark}, {et~al\mbox{.}}} \bibinfo{year}{2021}\natexlab{}.
\newblock \showarticletitle{Learning transferable visual models from natural language supervision}. In \bibinfo{booktitle}{\emph{International conference on machine learning}}. PMLR, \bibinfo{pages}{8748--8763}.
\newblock


\bibitem[Radford et~al\mbox{.}(2018)]%
        {radford2018improving}
\bibfield{author}{\bibinfo{person}{Alec Radford}, \bibinfo{person}{Karthik Narasimhan}, \bibinfo{person}{Tim Salimans}, \bibinfo{person}{Ilya Sutskever}, {et~al\mbox{.}}} \bibinfo{year}{2018}\natexlab{}.
\newblock \showarticletitle{Improving language understanding by generative pre-training}.
\newblock  (\bibinfo{year}{2018}).
\newblock


\bibitem[Rendle(2010)]%
        {rendle2010factorization}
\bibfield{author}{\bibinfo{person}{Steffen Rendle}.} \bibinfo{year}{2010}\natexlab{}.
\newblock \showarticletitle{Factorization machines}. In \bibinfo{booktitle}{\emph{2010 IEEE International conference on data mining}}. IEEE, \bibinfo{pages}{995--1000}.
\newblock


\bibitem[Rendle et~al\mbox{.}(2012)]%
        {rendle2012bpr}
\bibfield{author}{\bibinfo{person}{Steffen Rendle}, \bibinfo{person}{Christoph Freudenthaler}, \bibinfo{person}{Zeno Gantner}, {and} \bibinfo{person}{Lars Schmidt-Thieme}.} \bibinfo{year}{2012}\natexlab{}.
\newblock \showarticletitle{BPR: Bayesian personalized ranking from implicit feedback}.
\newblock \bibinfo{journal}{\emph{arXiv preprint arXiv:1205.2618}} (\bibinfo{year}{2012}).
\newblock


\bibitem[Sarwar et~al\mbox{.}(2001)]%
        {sarwar2001item}
\bibfield{author}{\bibinfo{person}{Badrul Sarwar}, \bibinfo{person}{George Karypis}, \bibinfo{person}{Joseph Konstan}, {and} \bibinfo{person}{John Riedl}.} \bibinfo{year}{2001}\natexlab{}.
\newblock \showarticletitle{Item-based collaborative filtering recommendation algorithms}. In \bibinfo{booktitle}{\emph{Proceedings of the 10th international conference on World Wide Web}}. \bibinfo{pages}{285--295}.
\newblock


\bibitem[Simonyan and Zisserman(2014)]%
        {simonyan2014very}
\bibfield{author}{\bibinfo{person}{Karen Simonyan} {and} \bibinfo{person}{Andrew Zisserman}.} \bibinfo{year}{2014}\natexlab{}.
\newblock \showarticletitle{Very deep convolutional networks for large-scale image recognition}.
\newblock \bibinfo{journal}{\emph{arXiv preprint arXiv:1409.1556}} (\bibinfo{year}{2014}).
\newblock


\bibitem[Sutskever et~al\mbox{.}(2014)]%
        {sutskever2014sequence}
\bibfield{author}{\bibinfo{person}{Ilya Sutskever}, \bibinfo{person}{Oriol Vinyals}, {and} \bibinfo{person}{Quoc~V Le}.} \bibinfo{year}{2014}\natexlab{}.
\newblock \showarticletitle{Sequence to sequence learning with neural networks}.
\newblock \bibinfo{journal}{\emph{Advances in neural information processing systems}}  \bibinfo{volume}{27} (\bibinfo{year}{2014}).
\newblock


\bibitem[Szegedy et~al\mbox{.}(2015)]%
        {szegedy2015going}
\bibfield{author}{\bibinfo{person}{Christian Szegedy}, \bibinfo{person}{Wei Liu}, \bibinfo{person}{Yangqing Jia}, \bibinfo{person}{Pierre Sermanet}, \bibinfo{person}{Scott Reed}, \bibinfo{person}{Dragomir Anguelov}, \bibinfo{person}{Dumitru Erhan}, \bibinfo{person}{Vincent Vanhoucke}, {and} \bibinfo{person}{Andrew Rabinovich}.} \bibinfo{year}{2015}\natexlab{}.
\newblock \showarticletitle{Going deeper with convolutions}. In \bibinfo{booktitle}{\emph{Proceedings of the IEEE conference on computer vision and pattern recognition}}. \bibinfo{pages}{1--9}.
\newblock


\bibitem[Tao et~al\mbox{.}(2022)]%
        {tao2022self}
\bibfield{author}{\bibinfo{person}{Zhulin Tao}, \bibinfo{person}{Xiaohao Liu}, \bibinfo{person}{Yewei Xia}, \bibinfo{person}{Xiang Wang}, \bibinfo{person}{Lifang Yang}, \bibinfo{person}{Xianglin Huang}, {and} \bibinfo{person}{Tat-Seng Chua}.} \bibinfo{year}{2022}\natexlab{}.
\newblock \showarticletitle{Self-supervised learning for multimedia recommendation}.
\newblock \bibinfo{journal}{\emph{IEEE Transactions on Multimedia}} (\bibinfo{year}{2022}).
\newblock


\bibitem[Tautkute et~al\mbox{.}(2019)]%
        {tautkute2019deepstyle}
\bibfield{author}{\bibinfo{person}{Ivona Tautkute}, \bibinfo{person}{Tomasz Trzci{\'n}ski}, \bibinfo{person}{Aleksander~P Skorupa}, \bibinfo{person}{{\L}ukasz Brocki}, {and} \bibinfo{person}{Krzysztof Marasek}.} \bibinfo{year}{2019}\natexlab{}.
\newblock \showarticletitle{Deepstyle: Multimodal search engine for fashion and interior design}.
\newblock \bibinfo{journal}{\emph{IEEE Access}}  \bibinfo{volume}{7} (\bibinfo{year}{2019}), \bibinfo{pages}{84613--84628}.
\newblock


\bibitem[Vaswani et~al\mbox{.}(2017)]%
        {vaswani2017attention}
\bibfield{author}{\bibinfo{person}{Ashish Vaswani}, \bibinfo{person}{Noam Shazeer}, \bibinfo{person}{Niki Parmar}, \bibinfo{person}{Jakob Uszkoreit}, \bibinfo{person}{Llion Jones}, \bibinfo{person}{Aidan~N Gomez}, \bibinfo{person}{{\L}ukasz Kaiser}, {and} \bibinfo{person}{Illia Polosukhin}.} \bibinfo{year}{2017}\natexlab{}.
\newblock \showarticletitle{Attention is all you need}.
\newblock \bibinfo{journal}{\emph{Advances in neural information processing systems}}  \bibinfo{volume}{30} (\bibinfo{year}{2017}).
\newblock


\bibitem[Wang et~al\mbox{.}(2021)]%
        {wang2021pyramid}
\bibfield{author}{\bibinfo{person}{Wenhai Wang}, \bibinfo{person}{Enze Xie}, \bibinfo{person}{Xiang Li}, \bibinfo{person}{Deng-Ping Fan}, \bibinfo{person}{Kaitao Song}, \bibinfo{person}{Ding Liang}, \bibinfo{person}{Tong Lu}, \bibinfo{person}{Ping Luo}, {and} \bibinfo{person}{Ling Shao}.} \bibinfo{year}{2021}\natexlab{}.
\newblock \showarticletitle{Pyramid vision transformer: A versatile backbone for dense prediction without convolutions}. In \bibinfo{booktitle}{\emph{Proceedings of the IEEE/CVF international conference on computer vision}}. \bibinfo{pages}{568--578}.
\newblock


\bibitem[Xiao et~al\mbox{.}(2022)]%
        {xiao2022training}
\bibfield{author}{\bibinfo{person}{Shitao Xiao}, \bibinfo{person}{Zheng Liu}, \bibinfo{person}{Yingxia Shao}, \bibinfo{person}{Tao Di}, \bibinfo{person}{Bhuvan Middha}, \bibinfo{person}{Fangzhao Wu}, {and} \bibinfo{person}{Xing Xie}.} \bibinfo{year}{2022}\natexlab{}.
\newblock \showarticletitle{Training large-scale news recommenders with pretrained language models in the loop}. In \bibinfo{booktitle}{\emph{Proceedings of the 28th ACM SIGKDD Conference on Knowledge Discovery and Data Mining}}. \bibinfo{pages}{4215--4225}.
\newblock


\bibitem[Xu et~al\mbox{.}(2021)]%
        {xu2021e2e}
\bibfield{author}{\bibinfo{person}{Haiyang Xu}, \bibinfo{person}{Ming Yan}, \bibinfo{person}{Chenliang Li}, \bibinfo{person}{Bin Bi}, \bibinfo{person}{Songfang Huang}, \bibinfo{person}{Wenming Xiao}, {and} \bibinfo{person}{Fei Huang}.} \bibinfo{year}{2021}\natexlab{}.
\newblock \showarticletitle{E2E-VLP: end-to-end vision-language pre-training enhanced by visual learning}.
\newblock \bibinfo{journal}{\emph{arXiv preprint arXiv:2106.01804}} (\bibinfo{year}{2021}).
\newblock


\bibitem[Yang et~al\mbox{.}(2022)]%
        {yang2022gram}
\bibfield{author}{\bibinfo{person}{Yoonseok Yang}, \bibinfo{person}{Kyu~Seok Kim}, \bibinfo{person}{Minsam Kim}, {and} \bibinfo{person}{Juneyoung Park}.} \bibinfo{year}{2022}\natexlab{}.
\newblock \showarticletitle{GRAM: Fast Fine-tuning of Pre-trained Language Models for Content-based Collaborative Filtering}.
\newblock \bibinfo{journal}{\emph{arXiv preprint arXiv:2204.04179}} (\bibinfo{year}{2022}).
\newblock


\bibitem[Yuan et~al\mbox{.}(2023)]%
        {yuan2023go}
\bibfield{author}{\bibinfo{person}{Zheng Yuan}, \bibinfo{person}{Fajie Yuan}, \bibinfo{person}{Yu Song}, \bibinfo{person}{Youhua Li}, \bibinfo{person}{Junchen Fu}, \bibinfo{person}{Fei Yang}, \bibinfo{person}{Yunzhu Pan}, {and} \bibinfo{person}{Yongxin Ni}.} \bibinfo{year}{2023}\natexlab{}.
\newblock \showarticletitle{Where to go next for recommender systems? id-vs. modality-based recommender models revisited}.
\newblock \bibinfo{journal}{\emph{arXiv preprint arXiv:2303.13835}} (\bibinfo{year}{2023}).
\newblock


\bibitem[Zhang et~al\mbox{.}(2021)]%
        {zhang2021mining}
\bibfield{author}{\bibinfo{person}{Jinghao Zhang}, \bibinfo{person}{Yanqiao Zhu}, \bibinfo{person}{Qiang Liu}, \bibinfo{person}{Shu Wu}, \bibinfo{person}{Shuhui Wang}, {and} \bibinfo{person}{Liang Wang}.} \bibinfo{year}{2021}\natexlab{}.
\newblock \showarticletitle{Mining latent structures for multimedia recommendation}. In \bibinfo{booktitle}{\emph{Proceedings of the 29th ACM International Conference on Multimedia}}. \bibinfo{pages}{3872--3880}.
\newblock


\bibitem[Zhang et~al\mbox{.}(2023b)]%
        {zhang2023multimodal}
\bibfield{author}{\bibinfo{person}{Lingzi Zhang}, \bibinfo{person}{Xin Zhou}, {and} \bibinfo{person}{Zhiqi Shen}.} \bibinfo{year}{2023}\natexlab{b}.
\newblock \showarticletitle{Multimodal Pre-training Framework for Sequential Recommendation via Contrastive Learning}.
\newblock \bibinfo{journal}{\emph{arXiv preprint arXiv:2303.11879}} (\bibinfo{year}{2023}).
\newblock


\bibitem[Zhang et~al\mbox{.}(2019)]%
        {zhang2019deep}
\bibfield{author}{\bibinfo{person}{Shuai Zhang}, \bibinfo{person}{Lina Yao}, \bibinfo{person}{Aixin Sun}, {and} \bibinfo{person}{Yi Tay}.} \bibinfo{year}{2019}\natexlab{}.
\newblock \showarticletitle{Deep learning based recommender system: A survey and new perspectives}.
\newblock \bibinfo{journal}{\emph{ACM computing surveys (CSUR)}} \bibinfo{volume}{52}, \bibinfo{number}{1} (\bibinfo{year}{2019}), \bibinfo{pages}{1--38}.
\newblock


\bibitem[Zhang et~al\mbox{.}(2023a)]%
        {zhang2023language}
\bibfield{author}{\bibinfo{person}{Zhipeng Zhang}, \bibinfo{person}{Piao Tong}, \bibinfo{person}{Yingwei Ma}, \bibinfo{person}{Qiao Liu}, \bibinfo{person}{Xujiang Liu}, {and} \bibinfo{person}{Xu Luo}.} \bibinfo{year}{2023}\natexlab{a}.
\newblock \showarticletitle{Language-Enhanced Session-Based Recommendation with Decoupled Contrastive Learning}.
\newblock \bibinfo{journal}{\emph{arXiv preprint arXiv:2307.10650}} (\bibinfo{year}{2023}).
\newblock


\bibitem[Zhao et~al\mbox{.}(2023)]%
        {zhao2023embedding}
\bibfield{author}{\bibinfo{person}{Xiangyu Zhao}, \bibinfo{person}{Maolin Wang}, \bibinfo{person}{Xinjian Zhao}, \bibinfo{person}{Jiansheng Li}, \bibinfo{person}{Shucheng Zhou}, \bibinfo{person}{Dawei Yin}, \bibinfo{person}{Qing Li}, \bibinfo{person}{Jiliang Tang}, {and} \bibinfo{person}{Ruocheng Guo}.} \bibinfo{year}{2023}\natexlab{}.
\newblock \showarticletitle{Embedding in Recommender Systems: A Survey}.
\newblock \bibinfo{journal}{\emph{arXiv preprint arXiv:2310.18608}} (\bibinfo{year}{2023}).
\newblock


\bibitem[Zhou et~al\mbox{.}(2018)]%
        {zhou2018deep}
\bibfield{author}{\bibinfo{person}{Guorui Zhou}, \bibinfo{person}{Xiaoqiang Zhu}, \bibinfo{person}{Chenru Song}, \bibinfo{person}{Ying Fan}, \bibinfo{person}{Han Zhu}, \bibinfo{person}{Xiao Ma}, \bibinfo{person}{Yanghui Yan}, \bibinfo{person}{Junqi Jin}, \bibinfo{person}{Han Li}, {and} \bibinfo{person}{Kun Gai}.} \bibinfo{year}{2018}\natexlab{}.
\newblock \showarticletitle{Deep interest network for click-through rate prediction}. In \bibinfo{booktitle}{\emph{Proceedings of the 24th ACM SIGKDD international conference on knowledge discovery \& data mining}}. \bibinfo{pages}{1059--1068}.
\newblock


\bibitem[Zhou et~al\mbox{.}(2023b)]%
        {zhou2023comprehensive}
\bibfield{author}{\bibinfo{person}{Hongyu Zhou}, \bibinfo{person}{Xin Zhou}, \bibinfo{person}{Zhiwei Zeng}, \bibinfo{person}{Lingzi Zhang}, {and} \bibinfo{person}{Zhiqi Shen}.} \bibinfo{year}{2023}\natexlab{b}.
\newblock \showarticletitle{A Comprehensive Survey on Multimodal Recommender Systems: Taxonomy, Evaluation, and Future Directions}.
\newblock \bibinfo{journal}{\emph{arXiv preprint arXiv:2302.04473}} (\bibinfo{year}{2023}).
\newblock


\bibitem[Zhou and Shen(2023)]%
        {zhou2023tale}
\bibfield{author}{\bibinfo{person}{Xin Zhou} {and} \bibinfo{person}{Zhiqi Shen}.} \bibinfo{year}{2023}\natexlab{}.
\newblock \showarticletitle{A tale of two graphs: Freezing and denoising graph structures for multimodal recommendation}. In \bibinfo{booktitle}{\emph{Proceedings of the 31st ACM International Conference on Multimedia}}. \bibinfo{pages}{935--943}.
\newblock


\bibitem[Zhou et~al\mbox{.}(2023a)]%
        {zhou2023bootstrap}
\bibfield{author}{\bibinfo{person}{Xin Zhou}, \bibinfo{person}{Hongyu Zhou}, \bibinfo{person}{Yong Liu}, \bibinfo{person}{Zhiwei Zeng}, \bibinfo{person}{Chunyan Miao}, \bibinfo{person}{Pengwei Wang}, \bibinfo{person}{Yuan You}, {and} \bibinfo{person}{Feijun Jiang}.} \bibinfo{year}{2023}\natexlab{a}.
\newblock \showarticletitle{Bootstrap latent representations for multi-modal recommendation}. In \bibinfo{booktitle}{\emph{Proceedings of the ACM Web Conference 2023}}. \bibinfo{pages}{845--854}.
\newblock


\end{thebibliography}

\appendix

\end{document}